\documentclass[10pt]{article}
\usepackage{graphicx}

\def\Title#1{\begin{center} {\Large #1 } \end{center}}
\def\Author#1{\begin{center}{ \sc #1} \end{center}}
\def\Address#1{\begin{center}{ \it #1} \end{center}}

\newcommand\pubblock{\rightline{\begin{tabular}{l} Proceedings of the Second Annual LHCP\\ \pubnumber\\
         \pubdate  \end{tabular}}}

\newenvironment{Abstract}{\begin{quotation} \begin{center} 
             \large ABSTRACT \end{center}\bigskip 
      \begin{center}\begin{large}}{\end{large}\end{center} \end{quotation}}

\newenvironment{Presented}{\begin{quotation} \begin{center} 
             PRESENTED AT\end{center}\bigskip 
      \begin{center}\begin{large}}{\end{large}\end{center} \end{quotation}}

\def\beq{\begin{equation}}
\def\eeq#1{\label{#1}\end{equation}}
\def\eeqn{\end{equation}}

\def\beqa{\begin{eqnarray}}
\def\eeqa#1{\label{#1}\end{eqnarray}}
\def\eeqan{\end{eqnarray}}

\let\bar=\overbar

\def\Dslash{\not{\hbox{\kern-4pt $D$}}}
\def\dslash{\not{\hbox{\kern-2pt $\del$}}}

\def\msb{{\bar{\ssstyle M \kern -1pt S}}}

\usepackage{color}
\usepackage{lineno}
\usepackage[colorlinks=true,linkcolor=blue,citecolor=blue,filecolor=black,urlcolor=blue]{hyperref}
\usepackage{amsmath}
\usepackage{amssymb}

\newcommand\pubnumber{ ATL-PHYS-PROC-2014-141 }
\newcommand\pubdate{\today}

\def\affiliation{
On behalf of the ATLAS, CMS, LHCb, CDF and D0 Experiments  \\
Physikalisches Institut \\ 
Universit\"at Bonn, Bonn 53115, Germany}

\begin{document}
%\linenumbers

% large size for the first page
\large
\begin{titlepage}
\pubblock

%% Change the title, name, abstract
%% Title 
\vfill
\Title{  BSM Higgs boson searches at LHC and the Tevatron  }
\vfill

%  if you need to add the support use this, fill the \support definition above. 
%   \Author{ FIRSTNAME LASTNAME \support }
\Author{ William Davey }
\Address{\affiliation}
\vfill
\begin{Abstract}

A review of the current experimental results on searches for Higgs bosons in
models beyond the Standard Model is presented. Searches from ATLAS, CMS and
LHCb use datasets from Run 1 of the LHC, including 7 and 8~TeV proton-proton
collisions. Searches from CDF and D0 use the full or partial 1.96~TeV
proton-anti-proton collision datasets from the Tevatron.

\end{Abstract}
\vfill

% DO NOT CHANGE 
\begin{Presented}
The Second Annual Conference\\
 on Large Hadron Collider Physics \\
Columbia University, New York, U.S.A \\ 
June 2-7, 2014
\end{Presented}
\vfill
\end{titlepage}
\def\thefootnote{\fnsymbol{footnote}}
\setcounter{footnote}{0}
%

% normal size for the rest
\normalsize

The discovery of a Higgs boson~\cite{ATLASHiggsDiscovery,CMSHiggsDiscovery} by
the ATLAS~\cite{ATLASDetector} and CMS~\cite{CMSDetector} experiments at the
Large Hadron Collider (LHC)~\cite{LHC} has shed light on the mechanism for
electroweak symmetry breaking. However, it still remains to be determined
whether the particle is indeed the predicted Higgs boson of the Standard Model
(SM), or whether it is something different --- a Higgs boson from Beyond the
Standard Model (BSM).  The consistency of the particle with the SM Higgs boson
has been demonstrated through its property measurements, which can also be used to
place indirect limits on BSM Higgs bosons~\cite{ATLASindirect,CMSindirect}.
However, many models with extended or modified Higgs sectors are also
consistent with these measurements, and the possibility of multiple Higgs
bosons still remains.  

%While measurements of the properties of the particle 
%are consistent with the SM, many models with extended or modified Higgs sectors 
%have not been ruled out and the possibility of multiple
%
%   Property measurements
%themselves can be used to place indirect limits on BSM Higgs bosons, and such
%limits have been published by ATLAS~\cite{ATLASindirect} and
%CMS~\cite{CMSindirect}. However, they cannot directly address 
%the possibility of multiple Higgs bosons. 

An alternate route to uncover the nature of the Higgs sector is via direct
searches.  This talk provides a review of searches for BSM Higgs bosons
conducted at the LHC and the Tevatron, including: 
%charged and neutral Higgs
%bosons within two-Higgs-doublet models (2HDMs) and the Minimal Supersymmetric
%Standard Model (MSSM), the Next-to-Minimal Supersymmetric Standard Model
%(NMSSM), cascade decays involving multiple Higgs bosons, Doubly charged Higgs
%bosons, Exotic decays of Higgs bosons into invisible or long-lived particles,
%and Fermiophobic Higgs bosons.
%
neutral and charged scalars within Two-Higgs-Doublet-Models (2HDMs), the
Minimal Supersymmetric Standard Model (MSSM) and the Next-To-Minimal
Supersymmetric Standard Model (NMSSM); doubly-charged scalars; cascade decays;
exotic decays to invisible or long-lived particles; and Fermiophobic models.

%\section{Searches for charged and neutral MSSM/2HDM Higgs bosons}
%\section{2HDMs, the MSSM and the NMSSM}
%\section{Extended Higgs sectors}

One of the simplest extensions to the Standard Model Higgs sector is an inclusion of
an additional Higgs doublet, resulting in a broad class of models known as
2HDMs. These models have a rich phenomenology, with five physical Higgs bosons
remaining after spontaneous symmetry breaking: two neutral $CP$-even ($h$ and
$H$), one neutral $CP$-odd ($A$) and two charged bosons ($H^{\pm}$).  The
MSSM is a particular case of a 2HDM in
which the whole Higgs sector can be described at tree-level by just two
additional parameters, typically chosen to be the ratio of the vacuum
expectation values of the two Higgs doublets, $\tan\beta$, and either the mass
of the $A$ or $H^{\pm}$ bosons, $m_A$ or $m_{H^{\pm}}$, respectively.
Couplings of the Higgs bosons to $\tau$-leptons and $b$-quarks can be
significantly enhanced, especially for large values of $\tan\beta$, making
searches in the $\tau\tau$ and $b\bar{b}$ modes the most sensitive to neutral
MSSM bosons.  The searches also typically take advantage of the strongly
enhanced $b$-quark associated production mechanism via the use of $b$-tagging.

\begin{figure}[htbp]
\centering
\includegraphics[width=0.45\textwidth]{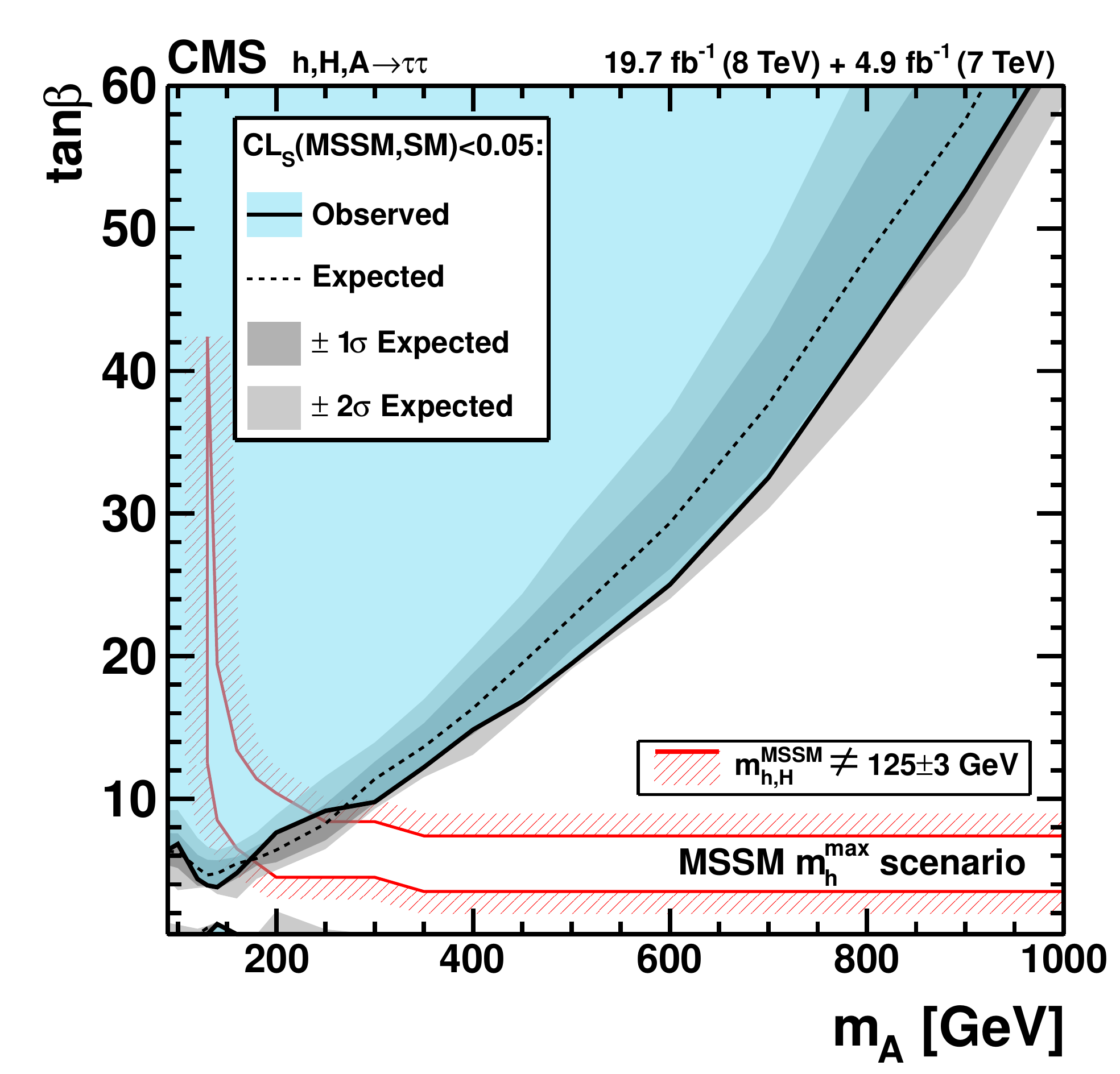} ~~
%\hspace{0.1\textwidth}
\includegraphics[width=0.45\textwidth]{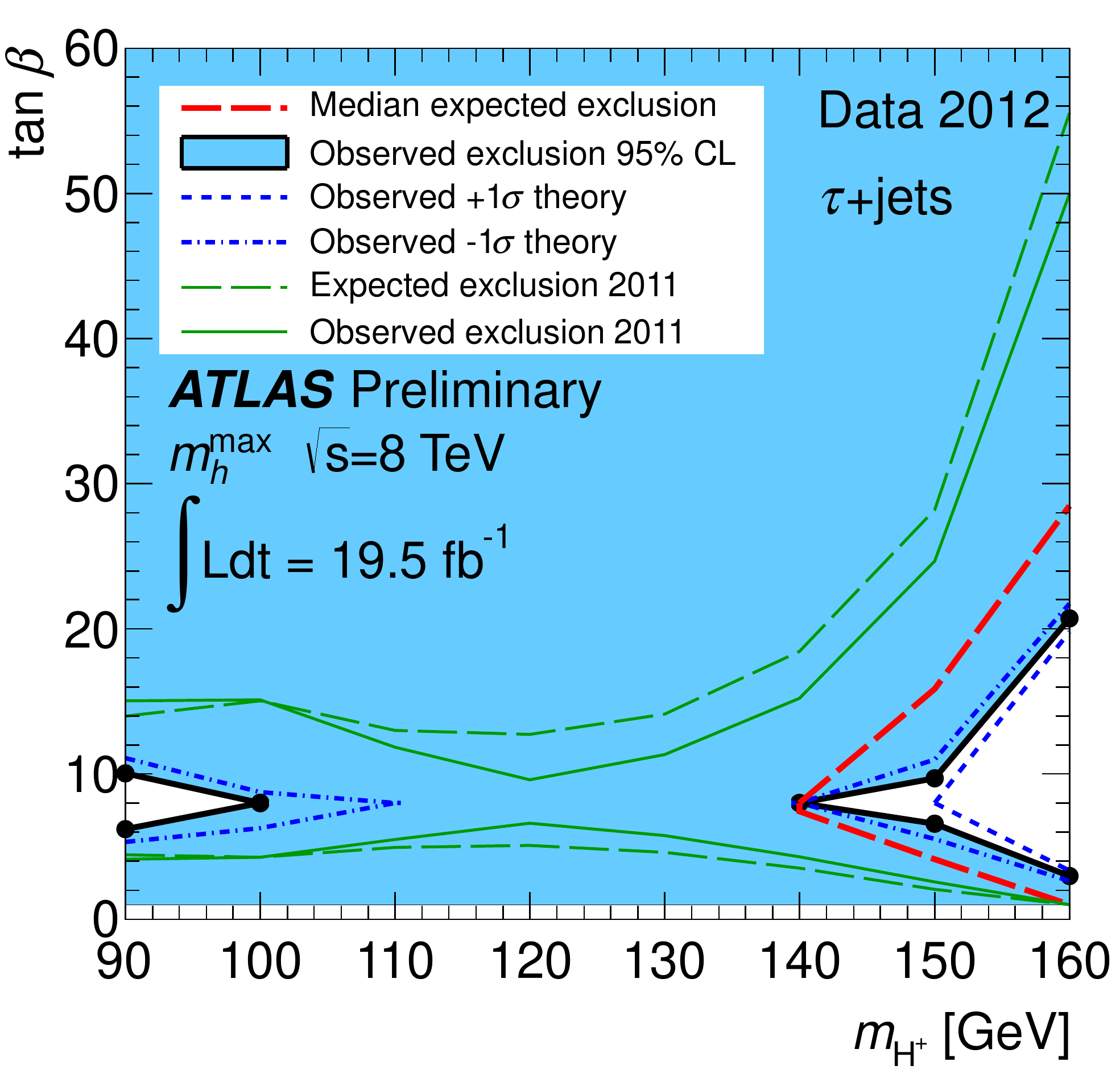}
\caption{
  Left: Limit in the MSSM parameter space from a search for $h/H/A\to\tau\tau$ by CMS~\cite{CMSMSSMHtautau}.
  Right: Limit in the MSSM parameter space from a search for $H\to\tau\nu$ by ATLAS~\cite{ATLASMSSMHtaunujets}.
}
\label{fig:MSSM}
\end{figure}

Searches for neutral Higgs bosons decaying to $\tau$-lepton pairs have been
performed by ATLAS~\cite{ATLASMSSMHtautau}, CMS~\cite{CMSMSSMHtautau},
LHCb~\cite{LHCbMSSMHtautau}, CDF~\cite{CDF2HDMHtautau} and D0~\cite{D0MSSMHtautaubb}. The most stringent
limits come from CMS, where the full 7 and 8 TeV datasets from the LHC are used.
The search analyses all possible $\tau\tau$ decay modes except the mode where
both $\tau$-leptons decay to $e\nu\nu$, which suffers large backgrounds and has a small
branching fraction. Events containing well identified decay products of two
oppositely charged $\tau$-leptons are selected. Further selection based on the
kinematic properties of the $\tau$ decays is used to suppress $W/Z$+jets events.
Finally, the events are categorised by the presence or absence of at least one
$b$-tagged jet. An excess of events over the SM background is searched for in
the reconstructed $\tau\tau$ mass distribution.  The dominant backgrounds come
from $Z/\gamma^{*}\to\tau\tau$ production and from $W$+jets and multijet
production where one or two jets, respectively, are misidentified as electrons,
muons or hadronic $\tau$ decays.  The search excludes large portions of the
parameter space for $m_A$ in the range \mbox{90~--~1000~GeV}, as seen in
Figure~\ref{fig:MSSM} (left).  Searches for neutral MSSM Higgs bosons in the
$b\bar{b}$ channel have also been performed by CDF and D0~\cite{TevMSSMHbb}, and
by CMS~\cite{CMSMSSMHbb}. However, these searches are not as sensitive due to
the large background from irreducible $b\bar{b}$ production.

Searches for charged Higgs bosons at the LHC and Tevatron have thus far
primarily targeted $H^{\pm}$ with a mass lower than the top-quark mass. In this
case, the dominant production is in $t\bar{t}$ decays, where  one top-quark
decays via a charged Higgs boson and the other via a $W$ boson.  For
$\tan\beta\gtrsim 3$, the charged Higgs boson decays predominantly via
$H^{\pm}\to\tau\nu$, while for $\tan\beta<3$, the $H\to c\bar{s}$ decay can be
significant. Searches are typically divided based on whether the $W$-boson and
$\tau$-lepton decay hadronically or leptonically.  Searches have been performed
by ATLAS~\cite{ATLASMSSMHtaunujets,ATLASMSSMHtaunulepvio,ATLASMSSMHcsbar},
CMS~\cite{CMSMSSMHtaunu} and CDF~\cite{CDFMSSMHcsbar}, covering the
$\tau$+jets, $\tau$+lepton and $c\bar{s}$+lepton final states.  The strongest
limits come from the $\tau$+jets channel at ATLAS, which uses the full 8~TeV
dataset from the LHC. Events containing exactly one identified $\tau$ candidate,
at least four additional jets (at least one $b$-tagged), no leptons and large
missing transverse momentum are selected.  An excess
over the dominant background from SM $t\bar{t}$ production is searched for in
the distribution of the transverse-mass between the missing transverse momentum
and the $\tau$ candidate.  The search excludes almost the entire available
parameter space for $90<m_{H^{\pm}}<160$\,GeV, as shown in
Figure~\ref{fig:MSSM} (right).  The search also places limits for
$m_{H^{\pm}}>m_{t}$, although the sensitivity is significantly reduced due to
the dominant $H^{\pm}\to t\bar{b}$ decay mode.

\begin{figure}[h]
\centering
\includegraphics[width=0.49\textwidth]{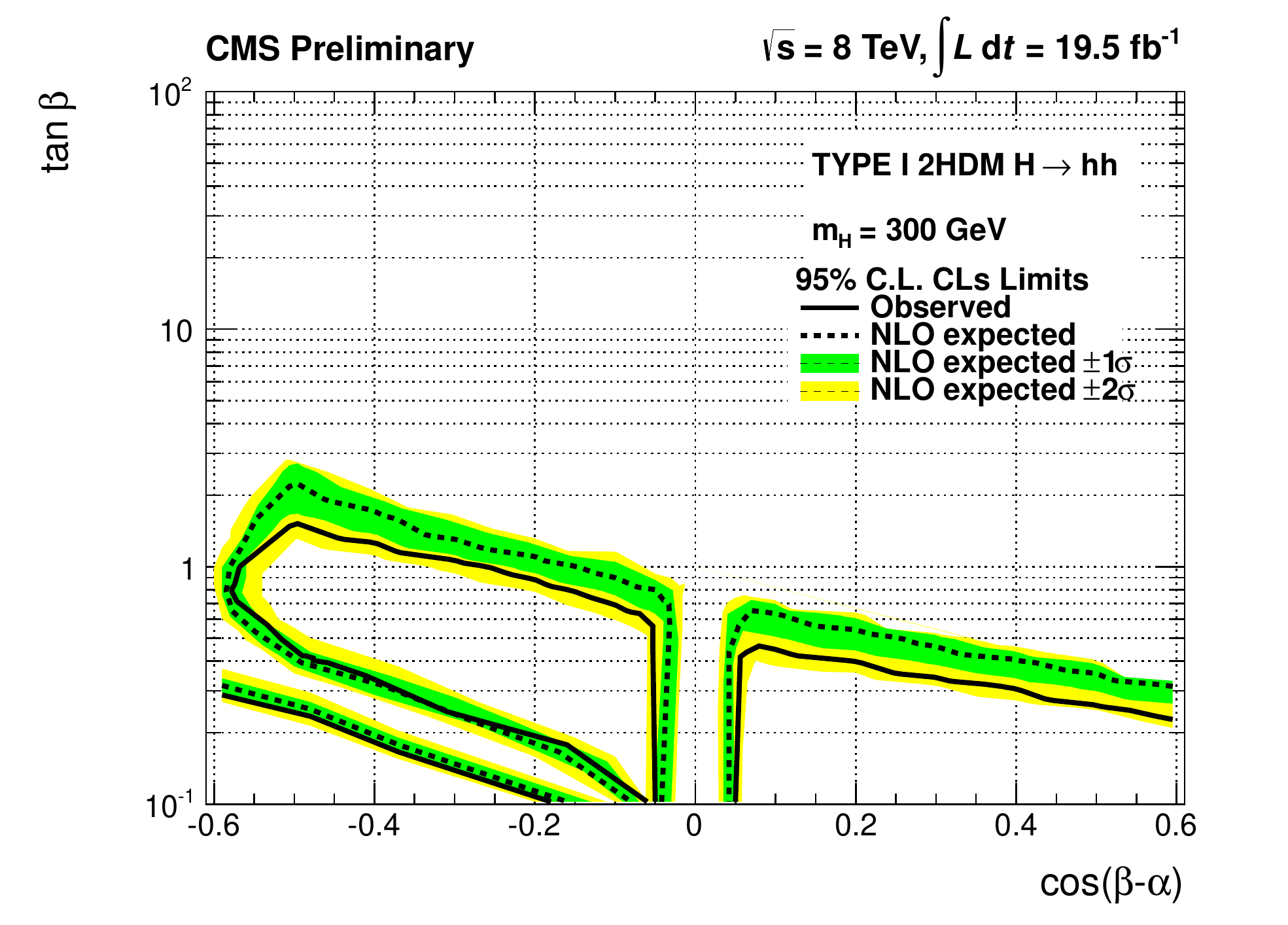}
\includegraphics[width=0.49\textwidth]{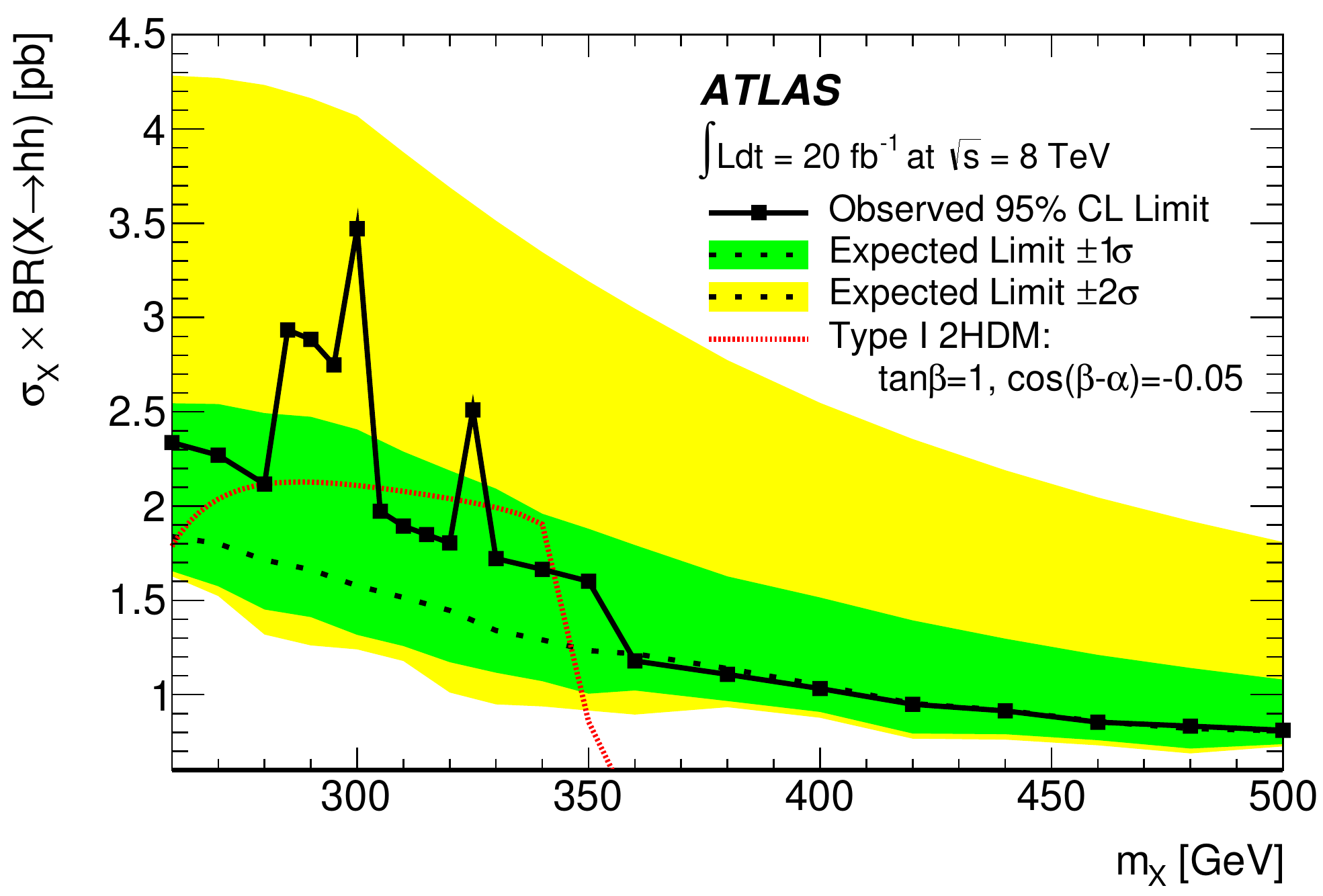}
\caption{
    Left: Limits on Type-I 2HDMs from a search for $H\to hh$ by
    CMS~\cite{CMSAZh}, where $\alpha$ is the $h$-$H$ mixing angle and $\tan\beta$
    is the ratio of the vacuum expectation values of the two Higgs doublets.
    Right: Limits on resonant $X\to hh \to \gamma\gamma b\bar{b}$ production from
    ATLAS~\cite{ATLASggbb}.
  } 
\label{fig:2HDM}
\end{figure}

2HDMs open up the possibility for a variety of {\em cascade decays} involving
multiple Higgs bosons. Searches for such decays have been performed by
ATLAS~\cite{ATLAS2HDMCascade} and CDF~\cite{CDF2HDMCascade} in the $H\to
W^-H^+(\to W^+h)$ channel, and CMS~\cite{CMSAZh} in the $H\to 2h$ and $A\to Zh$
channels. The signatures typically involve some combination of multi-leptons,
$b$-jets and photons.  The searches are able to exclude some regions of the
2HDM parameter space, as shown in Figure~\ref{fig:2HDM} (left) for Type-I
2HDMs~\cite{HiggsHunters}. Recently ATLAS has also published a search for
resonant and non-resonant production of two Higgs bosons in the $X\to hh \to
\gamma\gamma b\bar{b}$ mode~\cite{ATLASggbb}.  The corresponding limit for
resonant production is shown in Figure~\ref{fig:2HDM} (right).  Cascade decays
of Supersymmetric particles involving Higgs bosons are also
possible~\cite{SUSYPrimer}, but are not reviewed here. 

\begin{figure}[htb]
\centering
\includegraphics[width=0.58\textwidth]{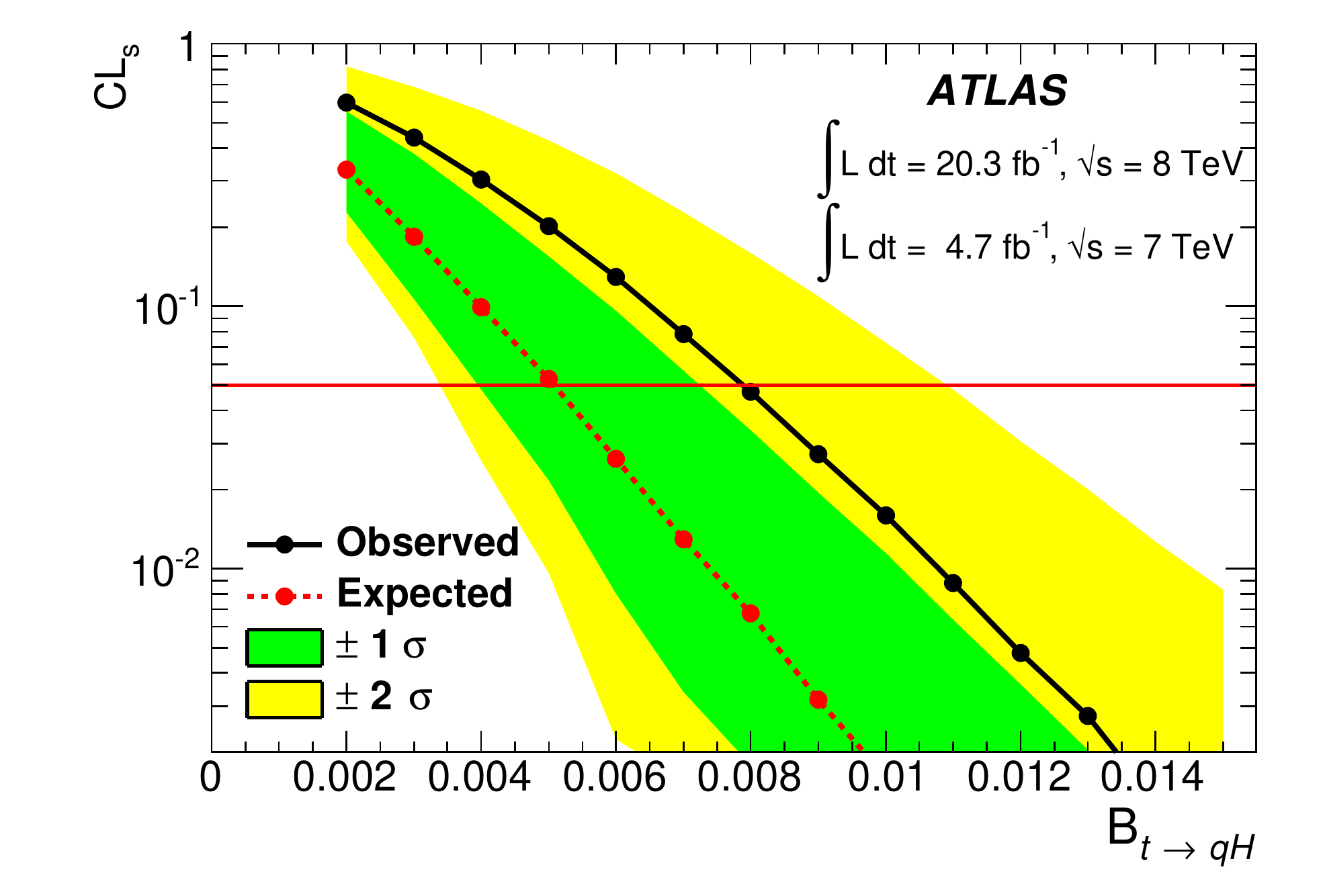}
\includegraphics[width=0.40\textwidth]{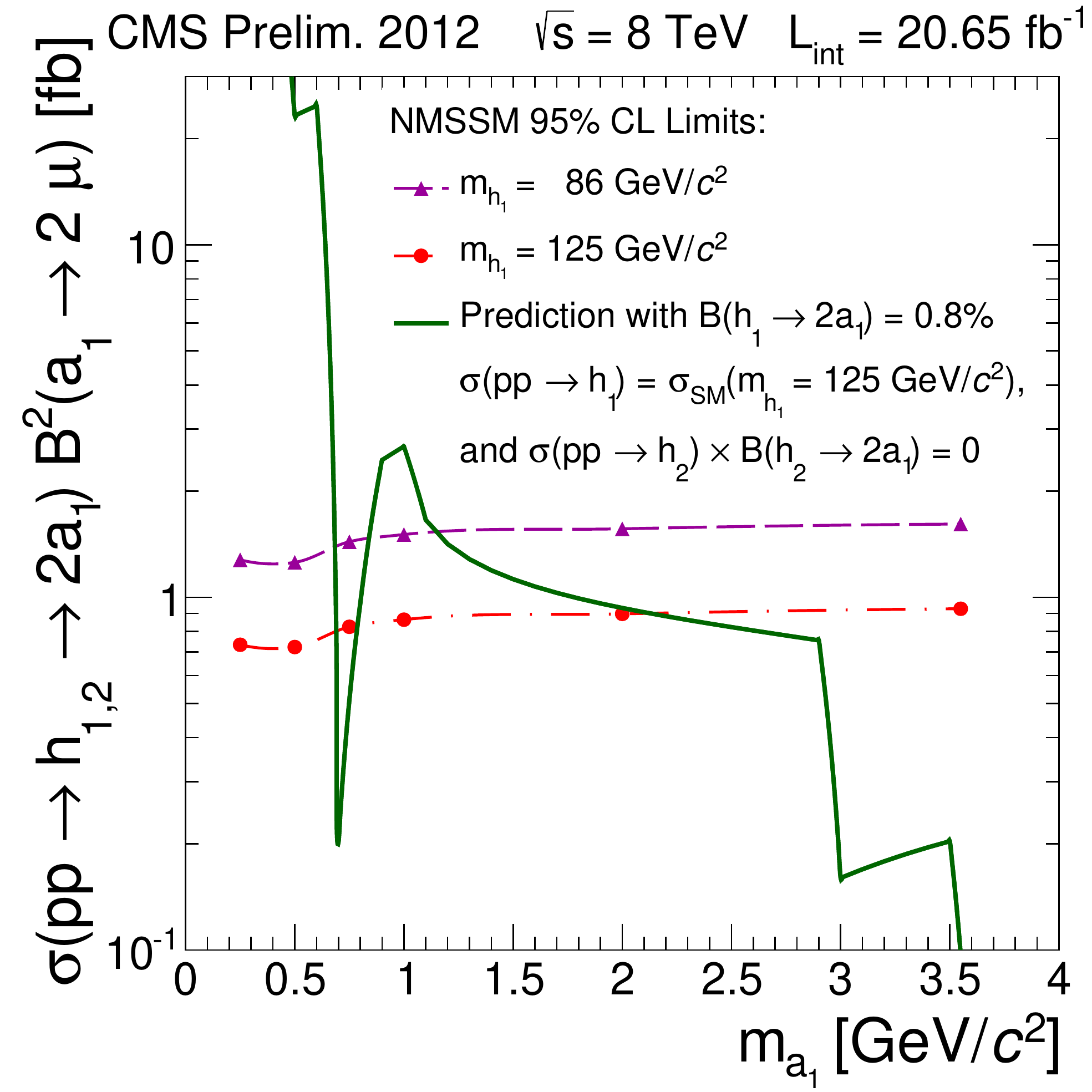}
\caption{
  Left: Limits on the flavour changing $t\to qH$ ($q=u,c$) process from
  ATLAS~\cite{ATLAStHc}.  Right: Limits on the lightest neutral Higgs boson of
  the NMSSM from a search for $h\to 2a \to 4\mu$ by CMS~\cite{CMSh2a4mu}.
}
\label{fig:THC}
\end{figure}

2HDMs can also exhibit large enhancements to processes involving flavour
changing neutral currents.  In particular, the $t\to Hc$ branching fraction,
which is predicted to be $3\cdot 10^{-15}$ in the SM can be as large as
$1.5\cdot 10^{-3}$ in some 2HDM models~\cite{2HDMIII}. Searches for the flavour
changing neutral current process $t\to Hc$ in $t\bar{t}$ events have been
performed by ATLAS~\cite{ATLAStHc} and CMS~\cite{CMStHc}.  In the search from
ATLAS, the $H\to\gamma\gamma$ decay is specifically targeted and a full
reconstruction of the $t\bar{t}$ system is performed, while the CMS result is a
reinterpretation of multilepton plus photon searches, which also covers $H\to
WW$ and $H\to ZZ$ modes. The searches put upper limits on ${\cal B}(t\to Hc)$
of 0.83\% and 0.56\%, respectively. The limit from ATLAS is shown in
Figure~\ref{fig:THC} (left).

In the NMSSM, a Higgs singlet is added to the MSSM Higgs sector. 
%, which alleviates the $\mu$ problem of the MSSM.
This results in seven physical Higgs bosons: five neutral and two charged.  In
the most general case, the neutral Higgs bosons can have mixed $CP$ states,
allowing for $CP$ violation in the Higgs sector. The lightest Higgs boson can
be very light (less than one GeV), in which case decays to $b\bar{b}$ are not
kinematically possible, which significantly alters the branching fractions.
Due to helicity suppression, the branching fraction to the heaviest possible
pair of particles is significantly enhanced. Decays into $\mu\mu$ are dominant
just above the dimuon production threshold, low-multiplicity states of light
hadrons become dominant above the $3\pi$ threshold, followed by $\tau\tau$ and
then $b\bar{b}$. Below the dimuon threshold, the $\gamma\gamma$ mode can be
sensitive.  Searches for a light neutral Higgs boson of the NMSSM have been
performed by ATLAS in the $a\to\mu\mu$~\cite{ATLASH2mu} and $h\to 2a\to
4\gamma$~\cite{ATLASH4gamma} channels, by CMS in the $a\to 2\mu$~\cite{CMSH2mu}
and $h\to 2a\to 4\mu$~\cite{CMSh2a4mu} channels, and by D0 in the $h\to 2a\to
4\mu$ and $h\to 2a\to 2\mu 2\tau$ channels~\cite{D0h2a2mu2tau}.  The limit from
the CMS search for $h\to 2a \to 4\mu$ is shown in Figure~\ref{fig:THC} (right).
The search is conducted by counting events that contain two well reconstructed
muon pairs with a good mass compatibility. The dominant backgrounds are from
$B$-meson and $J/\psi$ pair-production.  The search places strong limits for a
light Higgs boson in the mass range from $2m_{\mu}$ to $2m_{\tau}$.  The NMSSM also
opens up the possibility for decays of a charged Higgs boson via a light
neutral Higgs boson. CDF has performed a search for such a decay from a parent
top-quark: $t\to H^{+}b\to Wa(\to\tau\tau)b$~\cite{CDFChargedCascade}. They can
place limits as strong as ${\cal B}(t\to H^{+}b)<10\%$ for some regions of
parameter space.

\begin{figure}[htb]
\centering
\includegraphics[width=0.75\textwidth]{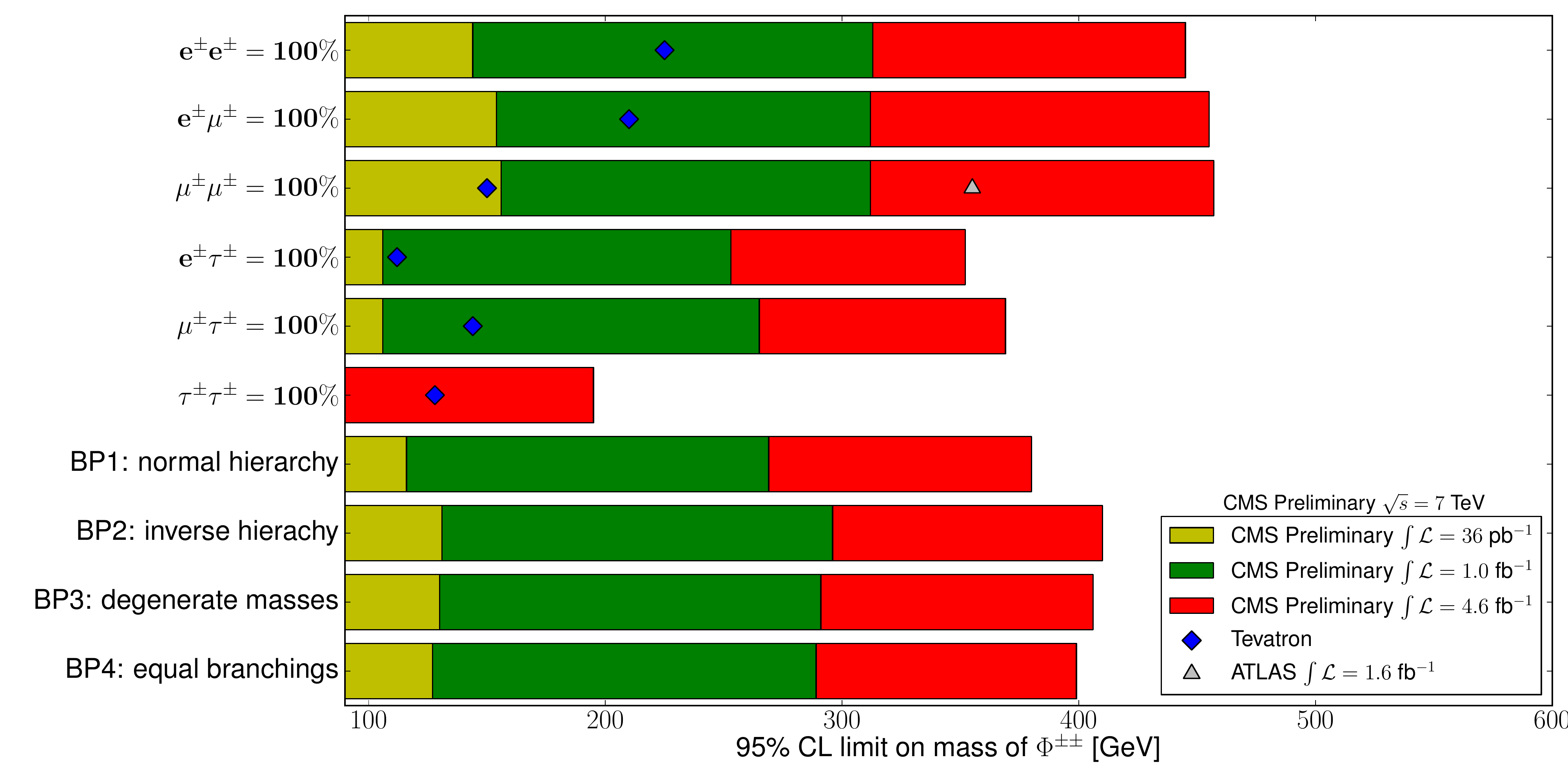}
%\hspace{0.1\textwidth}
%\includegraphics[width=0.33\textwidth]{CMSh4muLimit}
\caption{
  Limits on the mass of doubly charged Higgs bosons from CMS, ATLAS and Tevatron~\cite{CMSHpp}.
}
\label{fig:Hpp}
\end{figure}

Following the observation of non-zero neutrino masses through neutrino
oscillations, several mechanisms for neutrino mass generation have been
proposed, of which the see-saw mechanism is one of the most compelling.  In the
minimal type-II neutrino see-saw model, a scalar triplet is introduced, which
contains doubly charged Higgs bosons. Observation of such particles would
establish the type-II seesaw model as the most promising mechanism for neutrino
mass generation.  At the LHC and Tevatron, searches for doubly charged Higgs
bosons have been performed by ATLAS~\cite{ATLASHpp}, CMS~\cite{CMSHpp} and
D0~\cite{D0Hpp}. The searches typically require at least three reconstructed
leptons (including hadronic $\tau$ decays) and look for an excess of events
over the SM background in the mass distribution of same-sign lepton pairs.  The
searches put limits on the doubly charged Higgs boson mass of around 400\,GeV,
depending on the assumed branching fractions, as shown in Figure~\ref{fig:Hpp}. 

%\section{Exotic Higgs decays}
%\section{Other searches}

%% Invisible

\begin{figure}[htb]
\centering
\includegraphics[width=0.70\textwidth]{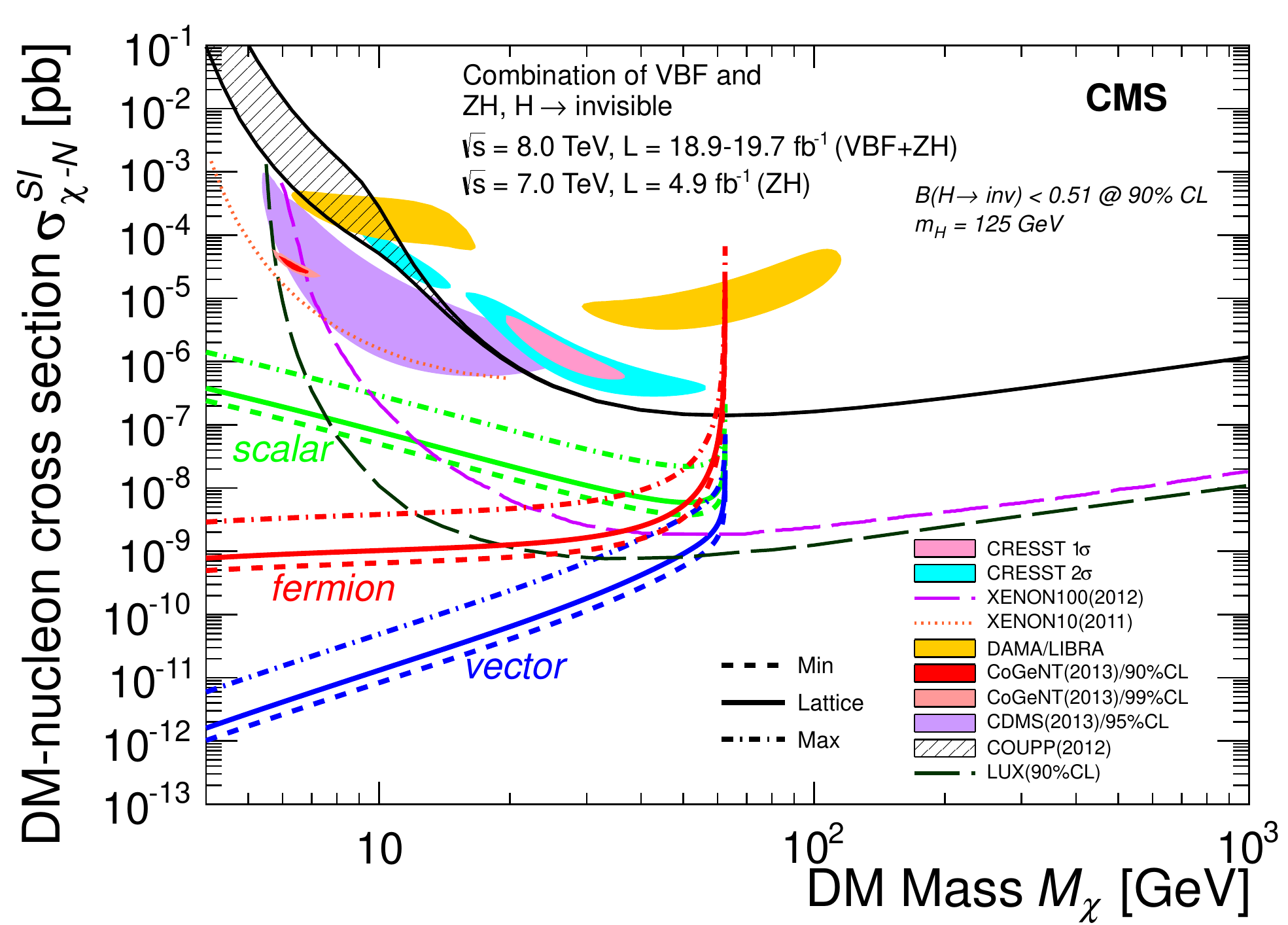}
\caption{ 
  Limits on {\em Higgs Portal} dark matter models from a combination of direct
  searches for Higgs decays to invisible particles
  % and measurements of Higgs
  %boson couplings to Standard Model particles 
  by CMS~\cite{CMSinvisible}.  The
  limits depend on the assumed nature of the dark-matter candidate: {\em vector},
  {\em fermion} or {\em scalar}. Indirect limits or preferred regions from direct
  detection experiments are overlaid.  
}
\label{fig:invisible}
\end{figure}

While coupling measurements provide constraints on the decay modes of the
discovered Higgs boson, the possibility of a non-negligible branching fraction
into exotic particles remains. Invisible decays of the Higgs boson are possible
in a wide range of models (e.g. neutralinos, graviscalars, etc.).  Searches for
invisible decays of the Higgs boson have been performed by
ATLAS~\cite{ATLASinvisible}, CMS~\cite{CMSinvisible} and
CDF~\cite{CDFinvisible}.  The searches rely on production of the Higgs boson in
association with a vector boson or via vector boson fusion, in which case the
momentum of the recoiling visible-system will lead to a large measured missing
transverse momentum.  
%The dominant backgrounds typically involve $Z\to\nu\nu$ decays in production
%with an additional vector boson or jets.  
The dominant backgrounds typically involve $Z\to\nu\nu$ or $W\to\ell\nu$ decays
in production with an additional $Z$ boson or jets.  The searches put an upper
limit of about 60\% at the 95\% confidence level on the invisible branching
fraction of the discovered Higgs boson. These limits
%, combined with constraints
%from measurements of the couplings of the Higgs boson to other SM particles,
can be used to place limits on {\em Higgs portal} models, in which the Higgs
boson provides an interaction between the SM and the dark-sector, as shown in
Figure~\ref{fig:invisible}. 

Some models, such as Hidden Valley models or Baryon Violating Supersymmetry,
predict decays of the Higgs boson to long-lived particles (e.g. v-hadrons or
the lightest Supersymmetric particle). The typical signature is
high-multiplicity jets of hadronic particles produced from a displaced vertex.
Searches for Higgs bosons decaying into exotic long-lived particles have been
performed by LHCb~\cite{LHCbLongLived}, D0~\cite{D0LongLived} and
CDF~\cite{CDFLongLived1,CDFLongLived2}.  The searches rely on the
reconstruction of at least 2 high-quality secondary vertices. The large
background due to vertices originating from an interaction with the detector
material is removed via geometrical selection. The dominant uncertainties
typically arise from the efficiency of the trigger and the vertex
reconstruction.  The searches have most sensitivity when the particles decay
far enough from the primary vertex so that they can be distinguished from the
hadronic background. However, sensitivity is lost quickly once the particles
decay outside the tracking volume. The search from D0 almost has enough
sensitivity to exclude the production of a SM-like Higgs boson decaying to a
pair of v-hadrons, depending on the properties of the v-hadrons.

Extensions of the SM Higgs sector, e.g. 2HDMs or Higgs triplet models, can also
predict substantially suppressed couplings to fermions. The most common way to
distinguish such models is to search for an enhancement in the
$H\to\gamma\gamma$ branching fraction, while $WW$ and $ZZ$ modes 
have also been considered.  Searches for {\em fermiophobic}
Higgs bosons have been performed by ATLAS~\cite{ATLASFermiophobic},
CMS~\cite{CMSFermiophobic}, CDF and D0~\cite{TevFermiophobic}. The CMS 
search excludes a purely fermiophobic Higgs boson with a mass in the 
range 110--147\,GeV. Slightly weaker limits are placed by the other 
experiments.

In conclusion, the LHC and Tevatron experiments have developed an extensive range of
searches for BSM Higgs bosons. While no significant deviations from the SM
have yet been observed, the stage is set for an exciting Run 2 of the LHC.

%% High-mass WW -> drop? 
%Searches for heavy Higgs bosons in the $WW$ channel have been performed by 
%ATLAS~\cite{ATLAShighmassWW,ATLAS2HDMWW} and CMS~\cite{CMShighmassWW}.

%\section{Conclusions}

%...... 

%%  if necessary
%\Acknowledgements
%I am grateful to XYZ for fruitful discussions.


\begin{thebibliography}{99}

%%
%%  bibliographic items can be constructed using the LaTeX format in SPIRES:
%%    see    http://www.slac.stanford.edu/spires/hep/latex.html
%%  SPIRES will also supply the CITATION line information; please include it.
%%

%% \bibitem{Aad:2012tfa} 
%%   G.~Aad {\it et al.}  [ATLAS Collaboration],
%%   %``Observation of a new particle in the search for the Standard Model Higgs boson with the ATLAS detector at the LHC,''
%%   Phys.\ Lett.\ B {\bf 716}, 1 (2012)
%%   [arXiv:1207.7214 [hep-ex]].
%%   %%CITATION = ARXIV:1207.7214;%%
%%   %3009 citations counted in INSPIRE as of 22 Jul 2014
%%   
%%   
%% %\cite{Chatrchyan:2012ufa}
%% \bibitem{Chatrchyan:2012ufa} 
%%   S.~Chatrchyan {\it et al.}  [CMS Collaboration],
%%   %``Observation of a new boson at a mass of 125 GeV with the CMS experiment at the LHC,''
%%   Phys.\ Lett.\ B {\bf 716}, 30 (2012)
%%   [arXiv:1207.7235 [hep-ex]].
%%   %%CITATION = ARXIV:1207.7235;%%
%%   %2951 citations counted in INSPIRE as of 22 Jul 2014



%%%%%%%%%%%%%%%%%%%%%%%%%%%%%%%%%%%%%%%%
%%
%% Starting the real bib items
%%
%%

%% -- Intro --

%% ATLASHiggsDiscovery
\bibitem{ATLASHiggsDiscovery} 
  ATLAS Collaboration, 
  \href{http://www.sciencedirect.com/science/article/pii/S037026931200857X}{Phys.\ Lett.\ B {\bf 716}, 1 (2012)}, 
  \href{http://arxiv.org/abs/1207.7214}{arXiv:1207.7214 [hep-ex]}.

%% CMSHiggsDiscovery
\bibitem{CMSHiggsDiscovery} 
  CMS Collaboration,
  \href{http://www.sciencedirect.com/science/article/pii/S0370269312008581}{Phys.\ Lett.\ B {\bf 716}, 30 (2012)},
  \href{http://arxiv.org/abs/1207.7235}{arXiv:1207.7235 [hep-ex]}.


%% ATLASDetector
\bibitem{ATLASDetector}
  ATLAS Collaboration,
  \href{http://dx.doi.org/10.1088/1748-0221/3/08/S08003}{JINST {\bf 3}, S08003 (2008)}.

%% CMSDetector
\bibitem{CMSDetector}
  CMS Collaboration,
  \href{http://dx.doi.org/10.1088/1748-0221/3/08/S08004}{JINST {\bf 3}, S08004 (2008)}.

%% LHC
\bibitem{LHC}
  L. Evans and P. Bryant (editors),
  \href{http://dx.doi.org/10.1088/1748-0221/3/08/S08001}{JINST, {\bf 3}, S08001 (2008)}.


%% ATLASindirect
\bibitem{ATLASindirect}
ATLAS Collaboration, ATLAS-CONF-2014-010, 
%\url{http://atlas.web.cern.ch/Atlas/GROUPS/PHYSICS/CONFNOTES/ATLAS-CONF-2014-010/}.
\url{http://cds.cern.ch/record/1670531}.

%% CMSindirect
\bibitem{CMSindirect}
CMS Collaboration,
\href{http://link.springer.com/article/10.1007%2FJHEP06%282013%29081}{JHEP {\bf 06} 081 (2013)},
\href{http://arxiv.org/abs/1303.4571}{arXiv:1303.4571 [hep-ex]}.



%% -- Htautau/bb --

%% ATLASMSSMHtautau
\bibitem{ATLASMSSMHtautau}
  ATLAS Collaboration, 
  \href{http://link.springer.com/article/10.1007/JHEP02(2013)095}{JHEP {\bf 02} 095 (2013)}, 
  \href{http://arxiv.org/abs/1211.6956}{arXiv:1211.6956 [hep-ex]}.

%% CMSMSSMHtautau
\bibitem{CMSMSSMHtautau}
  CMS Collaboration, {\em submitted to JHEP}, \href{http://arxiv.org/abs/1408.3316}{arXiv:1408.3316 [hep-ex]}.

%% LHCbMSSMHtautau
\bibitem{LHCbMSSMHtautau}
  LHCb Collaboration, R.~Aaij {\it et al.},
  \href{http://link.springer.com/article/10.1007%2FJHEP05(2013)132}{JHEP {\bf 05} 132 (2013)},
  \href{http://arxiv.org/abs/1304.2591}{arXiv:1304.2591 [hep-ex]}.



\bibitem{CDF2HDMHtautau}
  CDF Collaboration, T.~Aaltonen {\it et al.}, 
  %``Search for Higgs bosons predicted in two-Higgs-doublet models via decays to tau lepton pairs in 1.96-TeV p anti-p collisions,''
  \href{http://journals.aps.org/prl/abstract/10.1103/PhysRevLett.103.201801}{Phys.\ Rev.\ Lett.\  {\bf 103}, 201801 (2009)},
  \href{http://arxiv.org/abs/0906.1014}{arXiv:0906.1014 [hep-ex]}.
  %%CITATION = ARXIV:0906.1014;%%
  %49 citations counted in INSPIRE as of 15 Sep 2014

%% D0MSSMHtautaubb
\bibitem{D0MSSMHtautaubb}
  D0 Collaboration, V.M. Abazov {\it et al.}, 
\href{http://www.sciencedirect.com/science/article/pii/S0370269312002857}{Phys. Lett. B {\bf 710} 569 (2012)},
\href{http://arxiv.org/abs/1112.5431}{arXiv:1112.5431 [hep-ex]}.

%% TevMSSMHbb
\bibitem{TevMSSMHbb}
  CDF \& D0 Collaborations, T. Aaltonen {\it et al.}, \href{http://journals.aps.org/prd/abstract/10.1103/PhysRevD.86.091101}{Phys. Rev. D {\bf 86} 091101 (2012)}, \href{http://arxiv.org/abs/1207.2757}{arXiv:1207.2757 [hep-ex]}.

%% CMSMSSMHbb
\bibitem{CMSMSSMHbb} 
  CMS Collaboration, 
  \href{http://www.sciencedirect.com/science/article/pii/S0370269313002967}{Phys.\ Lett.\ B {\bf 722} 207 (2013)}, 
  \href{http://arxiv.org/abs/1302.2892}{arXiv:1302.2892 [hep-ex]}.



%% -- Charged Higgs --

%% ATLASMSSMHtaunujets
\bibitem{ATLASMSSMHtaunujets}
  ATLAS Collaboration,
  %\href{http://atlas.web.cern.ch/Atlas/GROUPS/PHYSICS/CONFNOTES/ATLAS-CONF-2013-090/}{ATLAS-CONF-2013-090}.
  ATLAS-CONF-2013-090,
  \url{http://cds.cern.ch/record/1595533}. 

%% ATLASMSSMHtaunulepvio
\bibitem{ATLASMSSMHtaunulepvio}
ATLAS Collaboration,
\href{http://link.springer.com/article/10.1007/JHEP03(2013)076}{JHEP {\bf 03} 076 (2013)},
\href{http://arxiv.org/abs/1212.3572}{arXiv:1212.3572 [hep-ex]}.

%% ATLASMSSMHcsbar
\bibitem{ATLASMSSMHcsbar}
ATLAS Collaboration,
\href{http://dx.doi.org/10.1140/epjc/s10052-013-2465-z}{EPJC {\bf 73} 2465 (2013)},
\href{http://arxiv.org/abs/1302.3694}{arXiv:1302.3694 [hep-ex]}.

%% CMSMSSMHtaunu
\bibitem{CMSMSSMHtaunu}
CMS Collaboration,
%\href{http://twiki.cern.ch/twiki/bin/view/CMSPublic/Hig12052TWiki}{CMS-HIG-12-052}.
CMS-HIG-12-052,
\url{http://cds.cern.ch/record/1502246}.

%% CDFMSSMHcsbar
\bibitem{CDFMSSMHcsbar}
CDF Collaboration, T. Aaltonen {\it et al.}, 
\href{http://journals.aps.org/prl/abstract/10.1103/PhysRevLett.103.101803}{Phys. Rev. Lett. {\bf 103} 101803 (2009)},
\href{http://arxiv.org/abs/0907.1269}{arXiv:0907.1269 [hep-ex]}.



%% -- Cascade --

%% ATLAS2HDMCascade
\bibitem{ATLAS2HDMCascade}
ATLAS Collaboration,
\href{http://journals.aps.org/prd/abstract/10.1103/PhysRevD.89.032002}{Phys. Rev. D {\bf 89} 032002 (2014)},
\href{http://arxiv.org/abs/1312.1956}{arXiv:1312.1956 [hep-ex]}.

%% CDF2HDMCascade
\bibitem{CDF2HDMCascade}
CDF Collaboration, T. Aaltonen {\it et al.}, 
\href{http://journals.aps.org/prl/abstract/10.1103/PhysRevLett.110.121801}{Phys. Rev. Lett. {\bf 110} 121801 (2013)},
\href{http://arxiv.org/abs/1212.3837}{arXiv:1212.3837 [hep-ex]}.

%% CMSAZh
\bibitem{CMSAZh}
CMS Collaboration,
%\href{http://twiki.cern.ch/twiki/bin/view/CMSPublic/Hig13025TWiki}{CMS-HIG-13-025}.
CMS-HIG-13-025,
\url{http://cds.cern.ch/record/1637776}.


%% HiggsHunters
\bibitem{HiggsHunters} 
  J.~F.~Gunion, H.~E.~Haber, G.~L.~Kane and S.~Dawson,
  %``The Higgs Hunter's Guide,''
  Front.\ Phys.\  {\bf 80}, 1 (2000).
  %%CITATION = FRPHA,80,1;%%
  %298 citations counted in INSPIRE as of 07 Sep 2014



%% ATLASggbb
\bibitem{ATLASggbb}
ATLAS Collaboration,
\href{http://arxiv.org/abs/1406.5053}{arXiv:1406.5053 [hep-ex]}. 


%% SUSYPrimer
\bibitem{SUSYPrimer} 
  S.~P.~Martin,
\href{http://www.worldscientific.com/doi/abs/10.1142/9789814307505_0001}{Adv.\ Ser.\ Direct.\ High Energy Phys.\  {\bf 21}, 1 (2010)},
\href{http://arxiv.org/abs/hep-ph/9709356}{arXiv:hep-ph/9709356}.



%% -- thc --

\bibitem{2HDMIII} 
  T.~P.~Cheng and M.~Sher,
  %``Mass Matrix Ansatz and Flavor Nonconservation in Models with Multiple Higgs Doublets,''
  \href{http://journals.aps.org/prd/abstract/10.1103/PhysRevD.35.3484}{Phys.\ Rev.\ D {\bf 35}, 3484 (1987)}.
  %%CITATION = PHRVA,D35,3484;%%
  %419 citations counted in INSPIRE as of 07 Sep 2014

%% ATLAStHc
\bibitem{ATLAStHc}
ATLAS Collaboration,
\href{http://link.springer.com/article/10.1007/JHEP06(2014)008}{JHEP {\bf 06} 008 (2014)},
\href{http://arxiv.org/abs/1403.6293}{arXiv:1403.6293 [hep-ex]}.

%% CMStHc
\bibitem{CMStHc}
CMS Collaboration, 
%\href{http://twiki.cern.ch/twiki/bin/view/CMSPublic/Hig13034TWiki}{CMS-HIG-13-034}.
CMS-HIG-13-034,
\url{http://cds.cern.ch/record/1666526}.




%% -- NMSSM --

%% ATLASH2mu
\bibitem{ATLASH2mu}
ATLAS Collaboration, 
%\href{http://atlas.web.cern.ch/Atlas/GROUPS/PHYSICS/CONFNOTES/ATLAS-CONF-2011-020/}{ATLAS-CONF-2011-020}.
ATLAS-CONF-2011-020,
\url{http://cds.cern.ch/record/1336749}.

%% ATLASH4gamma
\bibitem{ATLASH4gamma}
ATLAS Collaboration, 
%\href{http://atlas.web.cern.ch/Atlas/GROUPS/PHYSICS/CONFNOTES/ATLAS-CONF-2012-079/}{ATLAS-CONF-2012-079}.
ATLAS-CONF-2012-079,
\url{http://cds.cern.ch/record/1460391}.


%% CMSH2mu
\bibitem{CMSH2mu}
CMS Collaboration,
\href{http://journals.aps.org/prl/abstract/10.1103/PhysRevLett.109.121801}{Phys. Rev. Lett. {\bf 109} 121801 (2012)},
\href{http://arxiv.org/abs/1206.6326}{arXiv:1206.6326 [hep-ex]}.

%% CMSh2a4mu
\bibitem{CMSh2a4mu}
CMS Collaboration, 
%\href{http://twiki.cern.ch/twiki/bin/view/CMSPublic/Hig13010TWiki}{CMS-HIG-13-010}.
CMS-HIG-13-010,
\url{http://cds.cern.ch/record/1563546}.

%% D0h2a2mu2tau
\bibitem{D0h2a2mu2tau}
D0 Collaboration, V.M. Abazov {\it et al.}, 
\href{http://journals.aps.org/prl/abstract/10.1103/PhysRevLett.103.061801}{Phys. Rev. Lett. {\bf 103} 061801 (2009)},
\href{http://arxiv.org/abs/0905.3381}{arXiv:0905.3381 [hep-ex]}.

%% CDFChargedCascade
\bibitem{CDFChargedCascade}
CDF Collaboration, 
%\href{http://www-cdf.fnal.gov/physics/new/top/2009/tprop/nMSSMhiggs/cdf10104_chargedHiggs.pdf}{CDF Note 10104}.
CDF Note 10104,
\url{http://www-cdf.fnal.gov/physics/new/top/2009/tprop/nMSSMhiggs}.



%% -- Doubly Charged --

%% ATLASHpp
\bibitem{ATLASHpp}
ATLAS Collaboration,
\href{http://journals.aps.org/prd/abstract/10.1103/PhysRevD.85.032004}{Phys. Rev. D {\bf 85} 032004 (2012)},
\href{http://arxiv.org/abs/1201.1091}{arXiv:1201.1091 [hep-ex]}.

%% CMSHpp
\bibitem{CMSHpp}
CMS Collaboration,
\href{http://link.springer.com/article/10.1140/epjc/s10052-012-2189-5}{Eur. Phys. J. C {\bf 72} 2189 (2012)},
\href{http://arxiv.org/abs/1207.2666}{arXiv:1207.2666 [hep-ex]}.

%% D0Hpp
\bibitem{D0Hpp}
D0 Collaboration, V. M. Abazov {\it et al.},
\href{http://journals.aps.org/prl/abstract/10.1103/PhysRevLett.108.021801}{Phys. Rev. Lett. {\bf 108} 021801 (2012)},
\href{http://arxiv.org/abs/1106.4250}{arXiv:1106.4250 [hep-ex]}.


%% -- Invisible --

%% ATLASinvisible
\bibitem{ATLASinvisible}
ATLAS Collaboration,
\href{http://journals.aps.org/prl/abstract/10.1103/PhysRevLett.112.201802}{Phys. Rev. Lett. {\bf 112} 201802 (2014)},
\href{http://arxiv.org/abs/1402.3244}{arXiv:1402.3244 [hep-ex]}.

%% CMSinvisible
\bibitem{CMSinvisible}
CMS Collaboration,
submitted to EPJC,
\href{http://arxiv.org/abs/1404.1344}{arXiv:1404.1344 [hep-ex]}.

%% CDFinvisible
\bibitem{CDFinvisible}
CDF Collaboration, 
%\href{http://www-cdf.fnal.gov/physics/exotic/r2a/20140206.higgsinv/higgs_inv.html}{CDF Note 11068}.
CDF Note 11068,
\url{http://www-cdf.fnal.gov/physics/exotic/r2a/20140206.higgsinv/higgs_inv.html}.



%% -- Long Lived --

%% LHCbLongLived
\bibitem{LHCbLongLived}
LHCb Collaboration, R.~Aaij {\it et al.}, 
%\href{http://cds.cern.ch/record/1434432}{LHCb-CONF-2012-014}.
LHCb-CONF-2012-014,
\url{http://cds.cern.ch/record/1434432}.

%% D0LongLived
\bibitem{D0LongLived}
D0 Collaboration, V. M. Abazov {\it et al.}, 
\href{http://journals.aps.org/prl/abstract/10.1103/PhysRevLett.103.071801}{Phys. Rev. Lett. {\bf 103} 071801 (2009)},
\href{http://arxiv.org/abs/0906.1787}{arXiv:0906.1787 [hep-ex]}.

%% CDFLongLived1
\bibitem{CDFLongLived1}
CDF Collaboration, 
%\href{http://www-cdf.fnal.gov/physics/new/hdg/Results_files/results/HV_Dec2010/}{CDF Note 10356}.
CDF Note 10356,
\url{http://www-cdf.fnal.gov/physics/new/hdg/Results_files/results/HV_Dec2010}.

%% CDFLongLived2
\bibitem{CDFLongLived2}
CDF Collaboration,
%\href{http://www-cdf.fnal.gov/physics/new/hdg/Results_files/results/WZLeps_110805/}{CDF Note 10526}.
CDF Note 10526,
\url{http://www-cdf.fnal.gov/physics/new/hdg/Results_files/results/WZLeps_110805}.



%% -- Fermiophobic -- 

%% ATLASFermiophobic
\bibitem{ATLASFermiophobic}
ATLAS Collaboration,
\href{http://link.springer.com/article/10.1140/epjc/s10052-012-2157-0}{Eur.\ Phys.\ J.\ C {\bf 72} 2157 (2012)},
\href{http://arxiv.org/abs/1205.0701}{arXiv:1205.0701 [hep-ex]}.

%% CMSFermiophobic
\bibitem{CMSFermiophobic}
CMS Collaboration,
\href{http://www.sciencedirect.com/science/article/pii/S0370269313005182}{Phys. Lett. B {\bf 725} 36 (2013)},
\href{http://arxiv.org/abs/1302.1764}{arXiv:1302.1764 [hep-ex]}.

%% TevFermiophobic
\bibitem{TevFermiophobic}
CDF \& D0 Collaborations, T. Aaltonen {\it et al.}, 
\href{http://journals.aps.org/prd/abstract/10.1103/PhysRevD.88.052014}{Phys. Rev. D {\bf 88} 052014 (2103)},
\href{http://arxiv.org/abs/1303.6346}{arXiv:1303.6346 [hep-ex]}.


% %% ATLAShighmassWW
% \bibitem{ATLAShighmassWW}
% ATLAS Collaboration,
% \href{http://atlas.web.cern.ch/Atlas/GROUPS/PHYSICS/CONFNOTES/ATLAS-CONF-2013-067/}{ATLAS-CONF-2013-067}.
% 
% %% ATLAS2HDMWW
% \bibitem{ATLAS2HDMWW}
% ATLAS Collaboration,
% \href{http://atlas.web.cern.ch/Atlas/GROUPS/PHYSICS/CONFNOTES/ATLAS-CONF-2013-027/}{ATLAS-CONF-2013-027}.
% 
% %% CMShighmassWW
% \bibitem{CMShighmassWW}
% CMS Collaboration,
% \href{http://link.springer.com/article/10.1140%2Fepjc%2Fs10052-013-2469-8}{Eur. Phys. J. C {\bf 73} 2469 (2013)},
% \href{http://arxiv.org/abs/1304.0213}{arXiv:1304.0213 [hep-ex]}.



\end{thebibliography}
\end{document}